\documentclass[12pt,english,aps,pre,showpacs,reprint,longbibliography]{revtex4-1}
\usepackage{mathptmx}
\usepackage[T1]{fontenc}
\usepackage[latin9]{inputenc}
\usepackage{amsmath}
\usepackage{amssymb}
\usepackage{graphicx}

\makeatletter
\def\vec#1{\boldsymbol{\mathrm{#1}}}

\makeatother

\usepackage{babel}
\begin{document}

\title{Metamorphosis of helical magnetorotational instability in the presence
of axial electric current}

\author{J\={a}nis Priede }

\affiliation{Applied Mathematics Research Centre, Coventry University, Coventry,
CV1 5FB, United Kingdom}

\email{J.Priede@coventry.ac.uk}

\begin{abstract}
\noindent This paper presents numerical linear stability analysis
of a cylindrical Taylor-Couette flow of liquid metal carrying axial
electric current in a generally helical external magnetic field. Axially
symmetric disturbances are considered in the inductionless approximation
corresponding to zero magnetic Prandtl number. Axial symmetry allows
us to reveal an entirely new electromagnetic instability. First, we
show that the electric current passing through the liquid can extend
the range of helical magnetorotational instability (HMRI) indefinitely
by transforming it into a purely electromagnetic instability. Two
different electromagnetic instability mechanisms are identified. The
first is an internal pinch-type instability, which is due to the interaction
of the electric current with its own magnetic field. Axisymmetric
mode of this instability requires a free-space component of the azimuthal
magnetic field. When the azimuthal component of the magnetic field
is purely rotational and the axial component is nonzero, a new kind
of electromagnetic instability emerges. The latter, driven by the
interaction of electric current with a weak collinear magnetic field
in a quiescent fluid, gives rise to a steady meridional circulation
coupled with azimuthal rotation.
\end{abstract}


\maketitle

\section{Introduction}

Certain hydrodynamically stable rotational flows of electrically conducting
fluids can turn unstable in the presence of the magnetic field. This
rather counterintuitive effect was first predicted by \citet{Velikhov1959}
and Chandrasekhar \citep{Chandrasekhar1960,Chandrasekhar1961} for
cylindrical Taylor-Couette (TC) flow of a perfectly conducting fluid
subject to axial magnetic field. After three decades of obscurity
the MRI was re-discovered by Balbus and Hawley who speculated that
it could account for the fast formation of stars by driving turbulent
transport of angular momentum in accretion disks \citep{Balbus1991}.
This hypothesis has spurred many theoretical and numerical studies
\citep{Balbus1998} as well as several attempts to reproduce the MRI
in the laboratory \citep{Ji2013}. Though there is little doubt about
the reality of MRI, which follows directly from classical fluid mechanics
and electrodynamics, a convincing experimental demonstration of this
effect is hindered by a serious technical issue. Like the magnetohydrodynamic
dynamo, the MRI requires the magnetic Reynolds number $\mathrm{Rm}\sim10.$
For common liquid metals, which are relatively poor conductors characterized
by low magnetic Prandtl numbers $\mathrm{Pm}\sim10^{-5}-10^{-6},$
this translates into a large hydrodynamic Reynolds number $\mathrm{Re}=\mathrm{Rm}/\mathrm{Pm}\sim10^{6}-10^{7}$
\citep{Goodman2002}. At this high Reynolds numbers most flows become
turbulent due to inherently hydrodynamic mechanisms independent of
the MRI.

A way to circumvent this technical issue was suggested by \citet{Hollerbach2005},
who found that the threshold of MRI in cylindrical TC flow drops to
$\mathrm{Re}\sim10^{3}$ when the imposed magnetic field is helical
rather than purely axial as for the standard MRI (SMRI). This helical
type of MRI (HMRI) turned out to be significantly weaker and much
more limited than the SMRI \citep{Liu2006}. Nevertheless, an instability
closely resembling the HMRI was shortly observed in the PROMISE experiment
\citep{Stefani2006}. Subsequent analysis revealed that this instability
has been observed slightly beyond the narrow range in which the existence
of HMRI is predicted by the ideal TC flow model \citep{Priede2009}.
This apparently small discrepancy between the theory and experiment
hides two major issues pertinent to the HMRI. First, due to the hydrodynamic
\citep{Stefani2008a} and electromagnetic \citep{Priede2009} end
effects, the real base flow, in which the HMRI is to be observed,
inevitably deviates from the ideal TC flow used by the underlying
theory. The end effects can be reduced to some degree, as in the modified
PROMISE experiment \citep{Stefani2009}, but they cannot be eliminated
completely. Although the end effects can be taken into account by
realistic numerical models, which can achieve a good agreement with
the experiment, this does not solve the main problem, which is the
identification of the HMRI. Namely, the HMRI is physically indistinguishable
from a magnetically modified hydrodynamic Taylor vortex flow. The
distinction between both is only theoretical and based on the hydrodynamic
stability limit. The latter is well defined only for ideal TC flow
but not for a realistic base flow affected by the end effects. It
is not obvious how to determine this stability limit for a real base
flow affected by both hydrodynamic and magnetic end effects. Neither
experiment nor direct numerical simulation is able to discriminate
between the HMRI and other possible hydromagnetic of instabilities. 

The second issue that makes the identification of the HMRI particularly
hard is the very short extension of this instability, especially its
self-sustained (absolute) mode, beyond the hydrodynamic stability
limit \citep{Priede2009}. It is the narrow confinement of the HMRI
behind the hydrodynamic stability limit which makes the exact location
of this limit so important for the identification of the HMRI. Besides
the identification problem, the short extension of the HMRI implies
a limited astrophysical relevance of this instability. Namely, though
the HMRI is able to destabilize certain centrifugally stable velocity
distributions, it does not reach up to the astrophysically relevant
Keplerian rotation profile \citep{Liu2006,Priede2011,Kirillov2012}. 

Recently, it was suggested by \citet{Kirillov2013} that the range
of HMRI can significantly be extended when the azimuthal magnetic
field component is allowed to have a nonzero rotation. This apparently
minor mathematical modification of the model has several far-reaching
physical consequences which are the main concern of the present paper.
First , a nonpotential azimuthal magnetic field physically means the
presence of axial electric current in the fluid which provides an
electromagnetic energy source in addition to the mechanical rotation.
As a result, instability can develop without the background flow and
thus, in principle, extend over an unlimited range of velocity profiles.
In this paper we show that there are two such instabilities which
appear in the presence of background electric current. The first is
the resistive mode of internal pinch-type instability which was originally
predicted by \citet{Michael1954} in ideally conducting Taylor-Couette
flow bounded by solid walls where it is expected to develop on the
Alfvén time scale \citep{Velikhov1959}. The second appears to be
a new type of resistive instability driven by the interaction of axial
electric current with a weak collinear external magnetic field.

The first type of instability presents a certain astrophysical interest
as it is thought to affect the stars containing toroidal magnetic
fields \citep{Vandakurov1972,Tayler1973}. Because the strong radial
stratification in stellar interiors makes this instability nearly
horizontal and thus significantly differ from other pinch-type instabilities
\citep{Tayler1957}, \citet{Spruit1999} termed it Tayler instability.
This term was later used in a much broader sense by \citet{Ruediger2007}
to refer to current-driven instabilities in homogenous fluids including
liquid metals which are highly resistive from astrophysical point
of view. Such a resistive instability was presumably observed in the
recent liquid-metal experiment by \citet{Seilmayer2012}.

Although axial magnetic field has been extensively studied as a means
of stabilization of the plasma pinch \citep{Shafranov1956,Newcomb1960,Holm1985},
its potentially destabilizing effect in the highly resistive liquids
bounded by solid walls seems to have been overlooked so far. The previous
studies of the pinch instability in resistive fluids have been limited
to the conventional case of deformable boundaries \citep{Tayler1960}.
In this case, axial magnetic field applied along a liquid metal jet
carrying electric current is known to cause a kink instability\citep{Murty1961}.

The paper is organized as follows. The problem is formulated in Sec.
\ref{sec:prob}. Numerical results for various magnetic field configurations
are presented in Sec. \ref{sec:res}. The paper is concluded with
a brief discussion and summary of results in Sec. \ref{sec:end}.

\section{\label{sec:prob}Formulation of the problem}

\begin{figure}
\begin{centering}
\includegraphics[width=0.5\columnwidth]{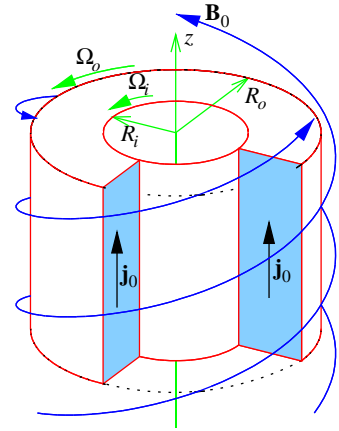}
\par\end{centering}

\protect\caption{\label{fig:sketch}(Color online) Sketch of the problem.}
\end{figure}

Consider an incompressible fluid of kinematic viscosity $\nu$ and
electrical conductivity $\sigma$ filling the gap between two infinite
concentric cylinders with the inner radius $R_{i}$ and the outer
radius $R_{o}$ rotating, respectively, with the angular velocities
$\Omega_{i}$ and $\Omega_{o}$ in the presence of a generally helical
magnetic field $\vec{B}_{0}=\vec{e}_{z}B_{z}+\vec{e}_{\phi}B_{\phi}$
with the axial component $B_{z}=\alpha B_{0}$ and the azimuthal component
\begin{equation}
B_{\phi}=B_{0}\left[(\beta-\gamma)R_{i}/r+\gamma r/R_{i}\right]\label{eq:b-phi}
\end{equation}
in cylindrical coordinates $(r,\phi,z).$ The dimensionless coefficient
$\alpha$ defines the magnitude of axial component of the magnetic
field relative to that of the azimuthal component. The latter has
a free-space part defined by the coefficient $\beta$ and a rotational
part defined by the coefficient $\gamma,$ which is associated with
the axial current density in the fluid $\vec{j}_{0}=\mu_{0}^{-1}\vec{\nabla}\times\vec{B}_{0}=\vec{e}_{z}\frac{2\gamma B_{0}}{\mu_{0}R_{i}},$
where $\mu_{0}$ is the magnetic permeability of vacuum. In the annular
geometry with $R_{i}\not=0,$ the absence of the current at $r<R_{i}$
produces also a free-space component of the magnetic field with the
effective helicity $-\gamma$ which appears in the first term of Eq.
(\ref{eq:b-phi}). Freespace magnetic field can be modified by passing
additional current along an electrode placed in the center of the
annular cavity as in the PROMISE experiment \citep{Stefani2006}.
This component of the magnetic field is specified by the coefficient
$\beta.$ Further we use $\alpha=1$ for the magnetic field with a
nonzero axial component, which means $B_{0}=B_{z}$ when $B_{z}\not=0$.
A purely azimuthal magnetic field corresponds to $\alpha=0.$

Following the inductionless approximation, which holds for most liquid-metal
magnetohydrodynamics characterized by small magnetic Reynolds numbers
$\mathrm{Rm}=\mu_{0}\sigma v_{0}L\ll1,$ where $v_{0}$ and $L$ are
the characteristic velocity and length scales, the magnetic field
of the currents induced by the fluid flow is assumed to be negligible
relative to the imposed field $\vec{B}_{0}$ everywhere except the
electromagnetic force term in the Navier-Stokes equation 

\begin{equation}
\partial_{t}\vec{v}+(\vec{v}\cdot\vec{\nabla})\vec{v}=\rho^{-1}\left(-\vec{\nabla}p+\vec{j}\times\vec{B}\right)+\nu\nabla^{2}\vec{v},\label{eq:N-S}
\end{equation}
where, as shown below, its interaction with the background electric
current $\vec{j}_{0}$ results in a non-negligible perturbation of
the electromagnetic body force. The electric current density is governed
by Ohm's law for a moving medium, 
\begin{equation}
\vec{j}=\sigma\left(\vec{E}+\vec{v}\times\vec{B}_{0}\right)\label{eq:Ohm}
\end{equation}
and related to the magnetic field by Ampère's law, $\vec{j}=\mu_{0}^{-1}\vec{\nabla}\times\vec{B}.$
In addition, we assume that the characteristic time of velocity variation
is much longer than the magnetic diffusion time $\tau_{0}\gg\tau_{m}=\mu_{0}\sigma L^{2}.$
This leads to the quasistationary approximation according to which
$\vec{\nabla}\times\vec{E}=0$ and $\vec{E}=-\vec{\nabla}\Phi,$ where
$\Phi$ is the electrostatic potential. Mass and charge conservation
imply $\vec{\nabla}\cdot\vec{v}=\vec{\nabla}\cdot\vec{j}=0.$

The problem admits a base state with a purely azimuthal velocity distribution
$\vec{v}_{0}(r)=\vec{e}_{\phi}v_{0}(r),$ where 
\[
v_{0}(r)=r\frac{\Omega_{o}R_{o}^{2}-\Omega_{i}R_{i}^{2}}{R_{o}^{2}-R_{i}^{2}}+\frac{1}{r}\frac{\Omega_{o}-\Omega_{i}}{R_{o}^{-2}-R_{i}^{-2}}.
\]
Note that this base flow is not affected by the magnetic field and
remains the same as in the hydrodynamic case. First, this is because
the unperturbed electromagnetic force is potential, and thus can be
compensated by a radial pressure gradient. Second, there is no current
and thus no additional electromagnetic force generated by the base
flow which gives rise only to the electrostatic potential $\Phi_{0}(r)=B_{0}\int v_{0}(r)\,dr,$
whose gradient compensates the induced electric field. Current can
appear only in the perturbed state 
\[
\left\{ \begin{array}{c}
\vec{v},p\\
\vec{B},\Phi
\end{array}\right\} (\vec{r},t)=\left\{ \begin{array}{c}
\vec{v}_{0},p_{0}\\
\vec{B}_{0},\Phi_{0}
\end{array}\right\} (r)+\left\{ \begin{array}{c}
\vec{v}_{1},p_{1}\\
\vec{B}_{1},\Phi_{1}
\end{array}\right\} (\vec{r},t),
\]
where $\vec{v}_{1},$ $p_{1},$ $\vec{B}_{1},$ and $\Phi_{1}$ are
small-amplitude perturbations for which Eqs. (\ref{eq:N-S} and \ref{eq:Ohm})
after linearization take the form 
\begin{eqnarray}
\partial_{t}\vec{v}_{1} & \negthickspace+\negthickspace & (\vec{v}_{1}\cdot\vec{\nabla})\vec{v}_{0}+(\vec{v}_{0}\cdot\vec{\nabla})\vec{v}_{1}\nonumber \\
 & = & \rho^{-1}\left(-\vec{\nabla}p_{1}+\vec{j}_{1}\times\vec{B}_{0}+\vec{j}_{0}\times\vec{B}_{1}\right)+\nu\nabla^{2}\vec{v}_{1}\label{eq:v1}\\
\vec{j}_{1} & = & \sigma\left(-\vec{\nabla}\Phi_{1}+\vec{v}_{1}\times\vec{B}_{0}\right)\,=\,\mu_{0}^{-1}\vec{\nabla}\times\vec{B}_{1}.\label{eq:j1}
\end{eqnarray}
Taking the \emph{curl} of Eq. (\ref{eq:j1}) to eliminate the potential
gradient we obtain the following induction equation
\begin{equation}
\sigma\vec{\nabla}\times\left(\vec{v}_{1}\times\vec{B}_{0}\right)+\mu_{0}^{-1}\nabla^{2}\vec{B}_{1}=0\label{eq:B1}
\end{equation}

The subsequent analysis is limited to axisymmetric perturbations which
are not necessary the most unstable but still useful for elucidating
the basic instability mechanisms. For such perturbations, the solenoidity
constraints are satisfied by introducing meridional stream functions
$\psi$ and $h$ for the fluid flow and electric current as 
\begin{eqnarray*}
\vec{v} & = & v\vec{e}_{\phi}+\vec{\nabla}\times(\psi\vec{e}_{\phi}),\\
\vec{j} & = & j\vec{e}_{\phi}+\vec{\nabla}\times(h\vec{e}_{\phi}).
\end{eqnarray*}
Note that $h$ is the azimuthal component of the induced magnetic
field which is governed by Eq. (\ref{eq:B1}) and used subsequently
instead of $\Phi$ for the description of the induced current. Equation
(\ref{eq:v1}) contains not only the azimuthal current, which is explicitly
related to the radial velocity, but also the radial component of the
induced magnetic field, which is subsequently denoted by $g$ and
governed by the radial component of Eq. (\ref{eq:B1}). For numerical
purposes, we introduce also the vorticity 
\[
\vec{\omega}=\omega\vec{e}_{\phi}+\vec{\nabla}\times(v\vec{e}_{\phi})=\vec{\nabla}\times\vec{v}
\]
 as an auxiliary variable. Perturbations are sought in the normal
mode form 
\[
\left\{ v_{1},\omega_{1,}\psi_{1},h_{1},g_{1}\right\} (\vec{r},t)=\left\{ \hat{v},\hat{\omega},\hat{\psi},\hat{h},\hat{g}\right\} (r)\times e^{\varGamma t+ikz},
\]
where $\varGamma$ is, in general, a generally complex growth rate
and $k$ is a real wave number. Henceforth, we proceed to dimensionless
variables by using $R_{i},$ $R_{i}^{2}/\nu,$ $R_{i}\Omega_{i},$
$B_{0},$ and $\sigma\mu_{0}B_{0}R_{i}^{2}\Omega_{i}$ as the length,
time, velocity, and the induced magnetic field scales, respectively.
Nondimensional governing equations then read as 
\begin{eqnarray}
\varGamma\hat{v} & = & D_{k}\hat{v}+\mathrm{Re}\,ikr^{-1}(r^{2}\Omega)'\hat{\psi}+\mathrm{Ha}^{2}(ik\alpha\hat{h}+2\gamma\hat{g}),\label{eq:vhat}\\
\varGamma\hat{\omega} & = & D_{k}\hat{\omega}+2\textit{Re}\,ik\Omega\hat{v}+\nonumber \\
 &  & \qquad\,+\,\textit{Ha}^{2}ik[ik\alpha\hat{\psi}-2((\beta-\gamma)r^{-2}+\gamma)\hat{h}],\label{eq:omghat}\\
0 & = & D_{k}\hat{\psi}+\hat{\omega},\label{eq:psihat}\\
0 & = & D_{k}\hat{h}+ik[\alpha\hat{v}-2(\beta-\gamma)r^{-2}\hat{\psi}],\label{eq:hhat}\\
0 & = & D_{k}\hat{g}+k^{2}\alpha\hat{\psi,}\label{eq:ghat}
\end{eqnarray}
 where $D_{k}f\equiv r^{-1}\left(rf'\right)'-(r^{-2}+k^{2})f$ and
the prime stands for $\frac{\mathrm{d}\,}{\mathrm{d}r};$ $\mathrm{Re}=R_{i}^{2}\Omega_{i}/\nu$
and $\mathrm{Ha}=R_{i}B_{0}\sqrt{\sigma/\rho\nu}$ are Reynolds and
Hartmann numbers, respectively; 
\[
\Omega(r)=\frac{\lambda^{-2}-\mu+r^{-2}\left(\mu-1\right)}{\lambda^{-2}-1}
\]
 is the dimensionless angular velocity of the base flow defined in
terms of $\lambda=R_{o}/R_{i}$ and $\mu=\Omega_{o}/\Omega_{i}$. 

\begin{figure*}
\begin{centering}
\includegraphics[width=0.45\textwidth]{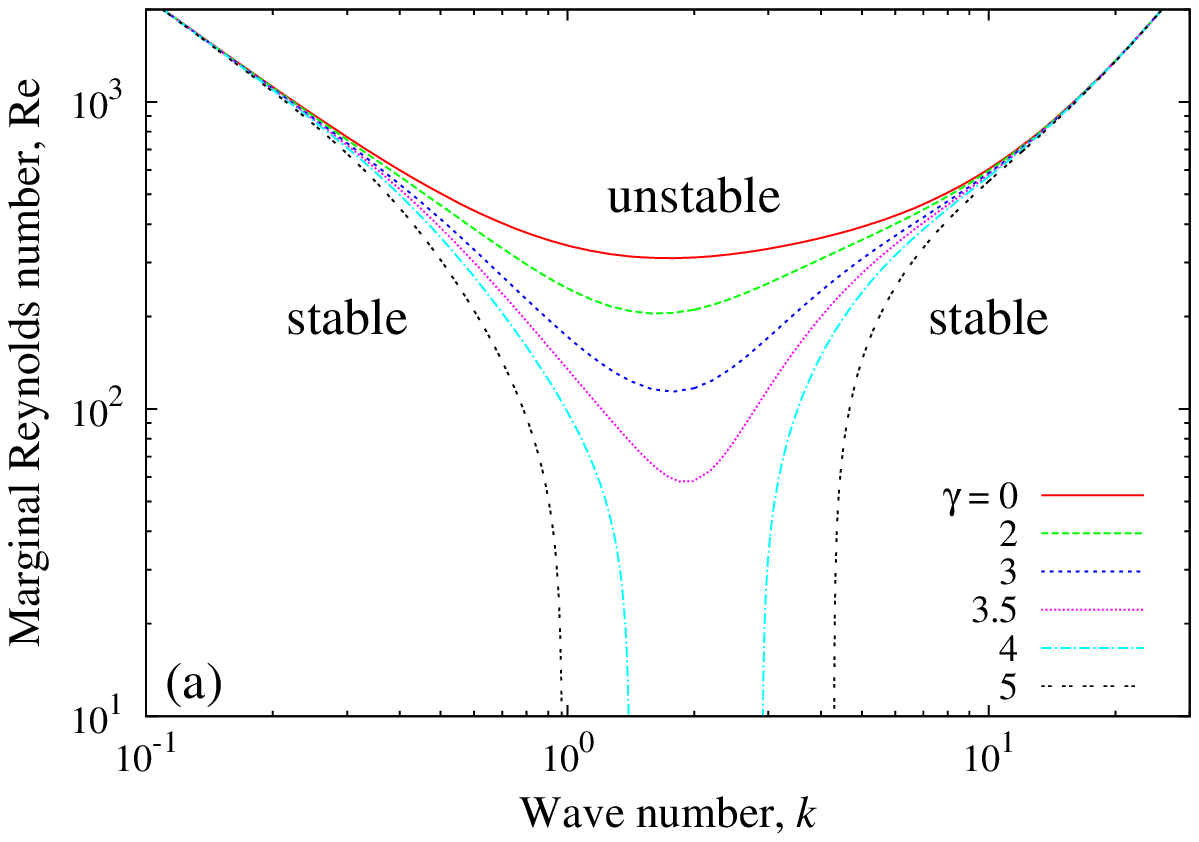}\includegraphics[width=0.45\textwidth]{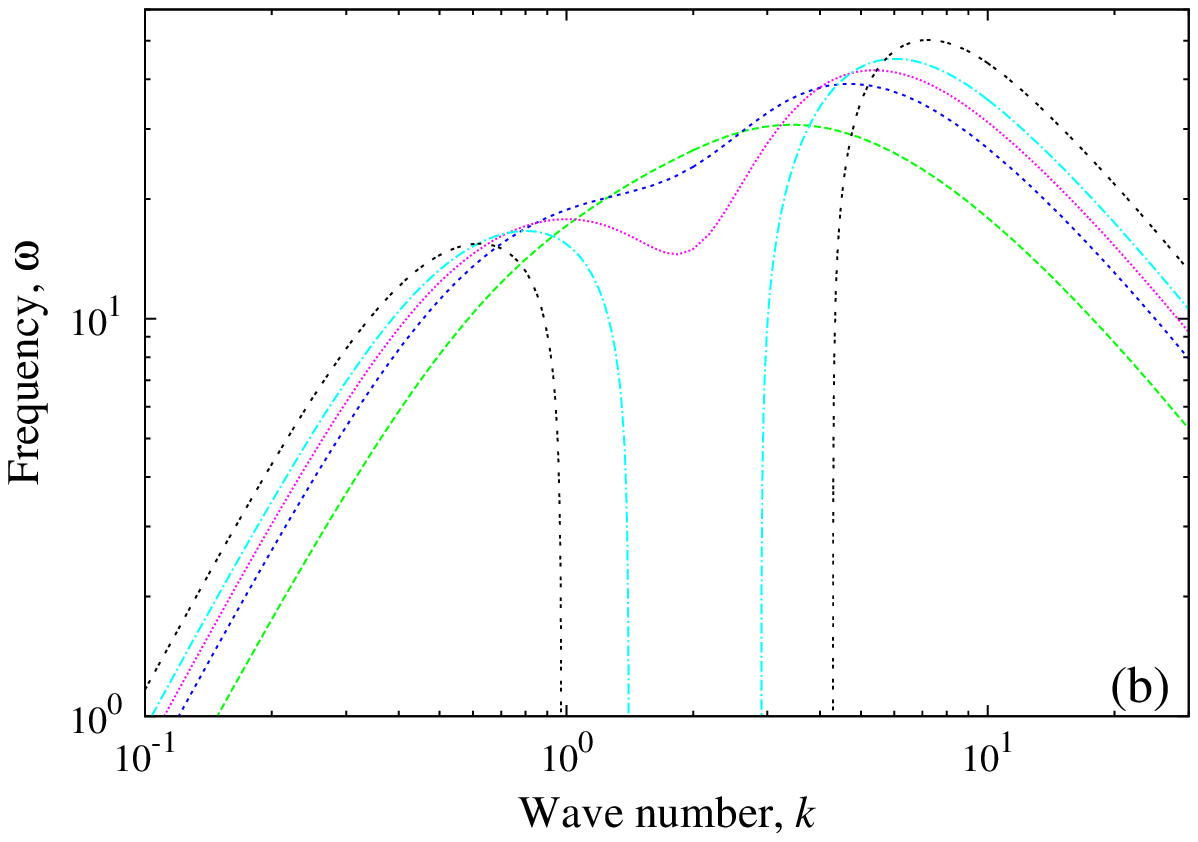}
\par\end{centering}

\protect\caption{\label{fig:rewk-mu0.2}(Color online) Marginal Reynolds number (a)
and the frequency (b) versus wave number for a hydrodynamically unstable
flow with $\mu=0.2$ at various helicities $\gamma$ of rotational
helical magnetic field with $\alpha=1,$ $\beta=0$ and $\mathrm{Ha}=10.$ }
\end{figure*}

The boundary conditions for the hydrodynamic perturbations on the
inner and outer cylinders at $r=1$ and $r=\lambda,$ respectively,
are $\hat{v}=\hat{\psi}=\hat{\psi}'=0.$ The boundary conditions for
the electric stream function $\hat{h}$ at insulating and perfectly
conducting cylinders are $\hat{h}=0$ and $(r\hat{h})'=0,$ respectively.
Note that the latter case should be understood as a limit only because
it implies an infinite current density in the perfectly conducting
walls in the presence of a non-zero axial electric current through
the liquid \citep{Edmonds1958}. The effective boundary conditions
for the radial component of the induced magnetic field $\hat{g}$
follow from the free-space solution of Eq. (\ref{eq:ghat}) with $\hat{\psi}\equiv0,$
which yields 
\[
\hat{g}(r)=\begin{cases}
G_{i}I_{1}(kr), & 0\le r\le1\\
G_{o}K_{1}(kr), & r\ge\lambda,
\end{cases}
\]
where $I_{1}$ and $K_{1}$ are the modified Bessel functions of the
first and second types of index 1 \citep{Abramowitz1964}. Taking
the ratio $(r\hat{g})'/\hat{g}$ to eliminate the unknown constants
$G_{i}$ and $G_{o}$, we obtain the sought boundary conditions in
the form 
\begin{eqnarray*}
(r\hat{g})'=c_{i}(kr)\hat{g} & \mbox{ at } & r=1,\\
(r\hat{g})'=c_{o}(kr)\hat{g} & \mbox{ at } & r=\lambda,
\end{eqnarray*}
where $c_{i}(r)=rI_{0}(r)/I_{1}(r)$ and $c_{o}(r)=-rK_{0}(r)/K_{1}(r).$
Because the radial magnetic field component $\hat{g}$ is generated
by the azimuthal current, which is tangential to the boundaries, it
is not affected by the conductivity of walls. Thus, the boundary conditions
above apply to both insulating and perfectly conducting cylinders.

Equations (\ref{eq:vhat})--(\ref{eq:ghat}) were solved numerically
using a spectral collocation method on a Chebyshev-Lobatto grid with
a typical number of internal points $N=32.$ In order to avoid spurious
eigenvalues, auxiliary Dirichlet boundary conditions for $\hat{\omega}$
were introduced and then numerically eliminated using the no-slip
boundary conditions $\hat{\psi}'=0$ \citep{Hagan2013}. The electromagnetic
variables $\hat{h}$ and $\hat{g}$ were represented in terms of $\hat{v}$
and $\hat{\psi}$ by numerical solution of Eqs. (\ref{eq:hhat} and
\ref{eq:ghat}) and then substituted into Eqs. (\ref{eq:vhat} and
\ref{eq:omghat}). The resulting standard complex matrix eigenvalue
problem of the size $2N\times2N$ was solved by the LAPACK ZGEEV routine.

\section{\label{sec:res}Results}

\subsection{Degeneration of the HMRI in the presence of axial electric current}

In the following, the radii ratio of inner and outer cylinders is
fixed to $\lambda=2$ and the cylinders are assumed to be insulating,
unless stated otherwise. We start with a hydrodynamically unstable
flow corresponding to the ratio of rotation rates $\mu=0.2,$ which
is below the Rayleigh limit $\mu_{c}=\lambda^{-2}=0.25.$ The magnetic
field is helical with the axial component fixed by $\alpha=1$ and
the azimuthal component generated only by the current passing through
the fluid, which corresponds to $\beta=0.$ In a purely axial magnetic
field corresponding to $\gamma=0,$ the flow becomes centrifugally
unstable to stationary Taylor vortices when Reynolds number exceeds
the marginal value which is plotted in Fig. \ref{fig:rewk-mu0.2}(a)
against the wave number $k$. Addition of a weak azimuthal magnetic
field reduces the instability threshold and makes the instability
oscillatory with the frequency $\omega=\Im[\varGamma${]} which is
shown in Fig. \ref{fig:rewk-mu0.2}(b). The most important result
seen in Fig. \ref{fig:rewk-mu0.2}(a) is the drop of marginal Reynolds
number to zero in a range of intermediate wave numbers when the helicity
of the field due the axial current defined by $\gamma$ becomes somewhat
greater than $3.7.$ Zero Reynolds number means that this instability
becomes entirely electromagnetic. Moreover, Fig. \ref{fig:rewk-mu0.2}(b)
shows that this instability is stationary, i.e., $\omega=0.$ It will
be shown later that two different electromagnetic mechanisms may be
behind this instability.

\begin{figure*}
\begin{centering}
\includegraphics[width=0.45\textwidth]{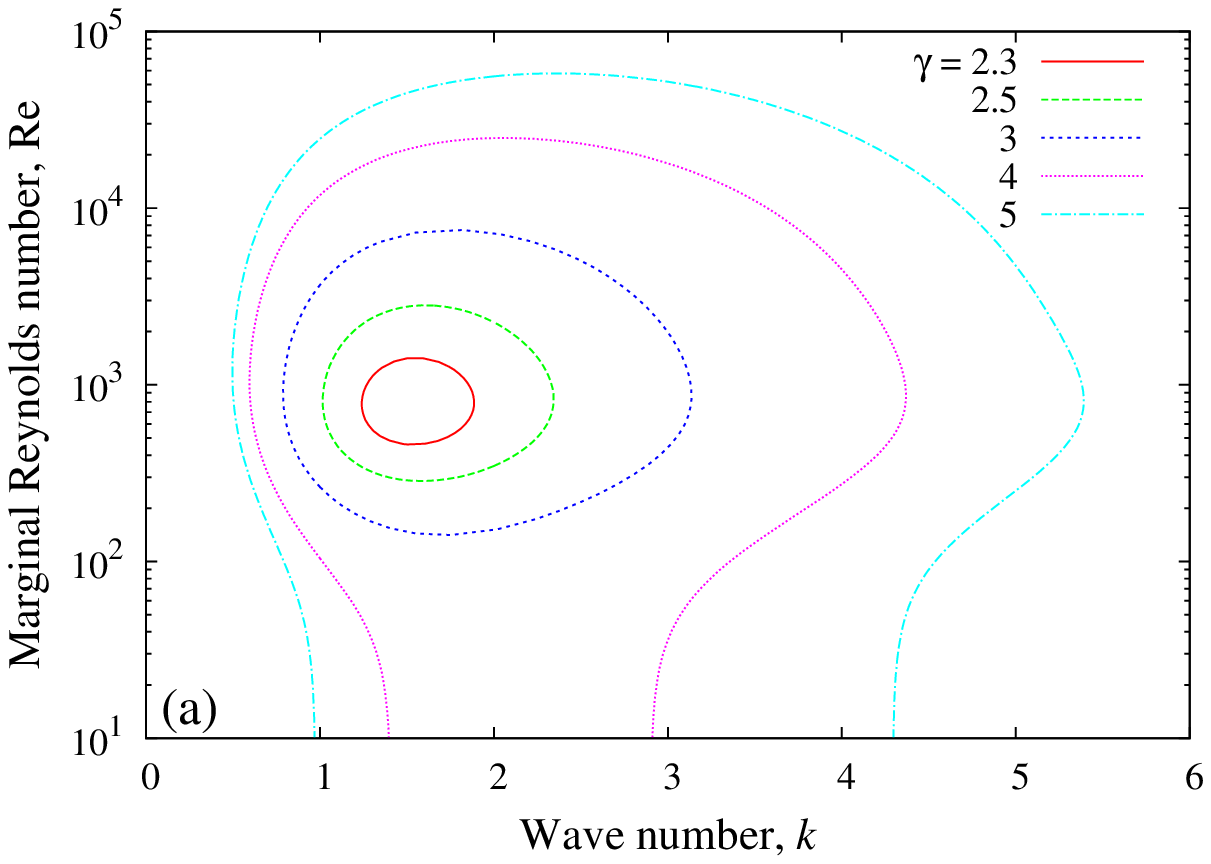}\includegraphics[width=0.45\textwidth]{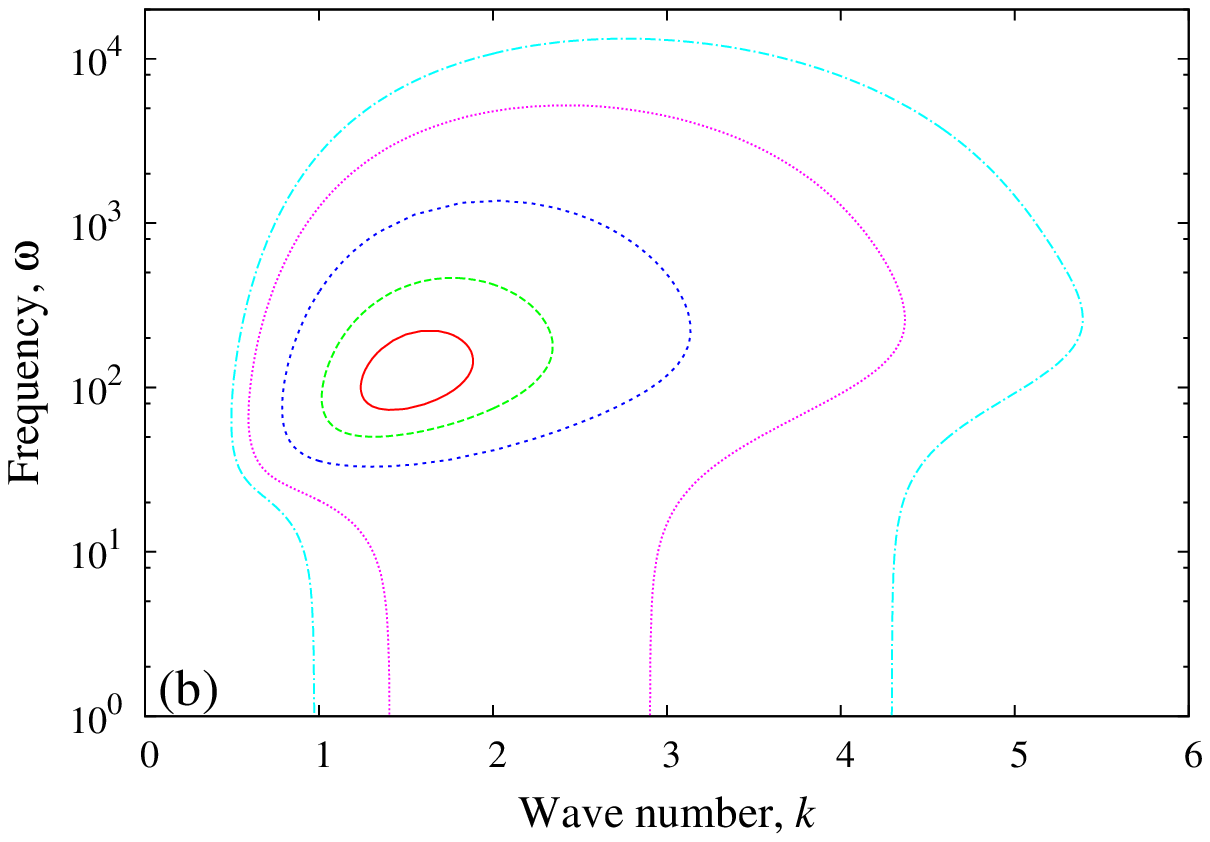}
\par\end{centering}

\protect\caption{\label{fig:rewk-mu0.3}(Color online) Marginal Reynolds number (a)
and the frequency (b) versus the wave number for a hydrodynamically
stable flow with $\mu=0.3$ at various helicities $\gamma$ of rotational
helical magnetic field with $\alpha=1,$ $\beta=0$, and $\mathrm{Ha}=10.$ }
\end{figure*}

Next, let us turn to a hydrodynamically stable case corresponding
to the ratio of rotation rates set to $\mu=0.3$, which is slightly
above the Rayleigh limit $\mu_{c}=0.25.$ As seen in Fig. \ref{fig:rewk-mu0.3}(a),
a moderately helical rotational magnetic field can destabilize this
flow similarly to the helical free-space magnetic field \citep{Priede2007}.
In both cases neutral stability curves form closed contours, which
means that the instability can occur only in limited ranges of Reynolds
and wave numbers. In contrast to the hydrodynamically unstable case
considered above, there are now two marginal Reynolds numbers - the
lower one by exceeding which the flow destabilizes, and the upper
one by exceeding which the flow restabilizes. The existence of the
upper critical Reynolds number is another peculiarity of the HMRI
which, in principle, distinguishes it from a magnetically modified
Taylor vortex flow \citep{Priede2007}. The upper critical Reynolds
number and the associated islands of instability appear also in a
centrifugally unstable regime with an axial through flow \citep{Altmeyer2011}.

\begin{figure*}
\begin{centering}
\includegraphics[width=0.45\textwidth]{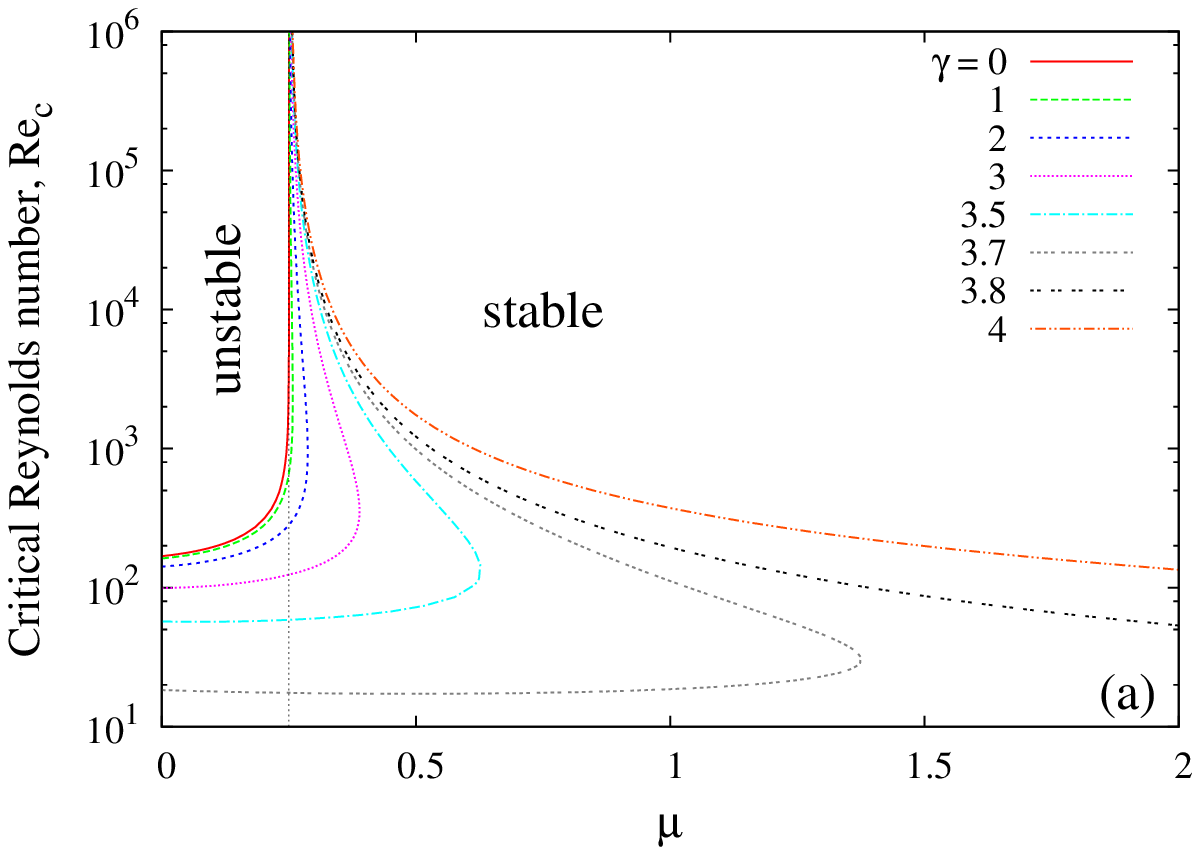}\includegraphics[width=0.45\textwidth]{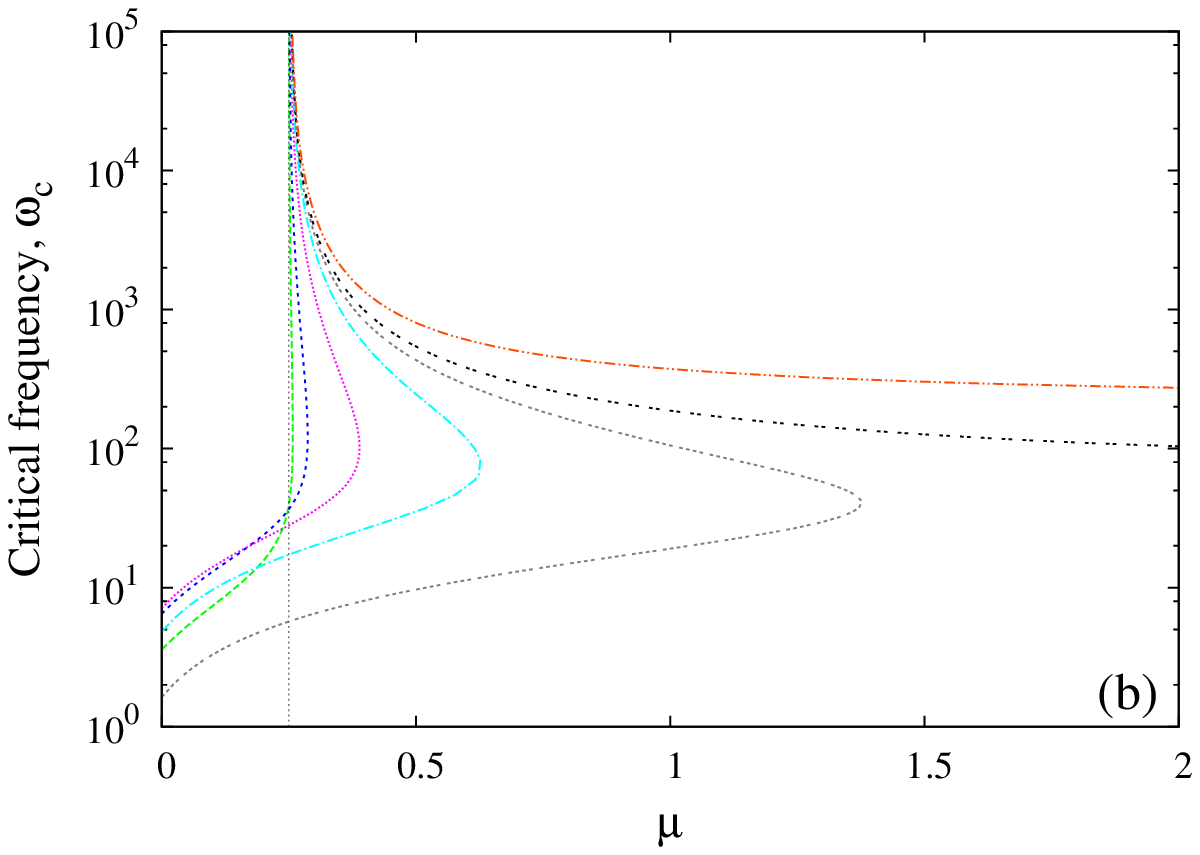}
\par\end{centering}

\protect\caption{\label{fig:rec-gm}(Color online) Critical Reynolds number (a) and
the frequency (b) versus the ratio of rotation rates of inner and
outer cylinders $\mu$ at various helicities $\gamma$ of rotational
helical magnetic field with $\alpha=1,$ $\beta=0$ and $\mathrm{Ha}=10.$ }
\end{figure*}

This picture changes when the helicity of the rotational field exceeds
$\gamma\approx3.7.$ As for the hydrodynamically unstable case considered
above, marginal Reynolds number again drops to zero in a certain range
of intermediate wave numbers. Figure \ref{fig:rec-gm}(a) shows the
critical Reynolds number and the respective frequency versus the ratio
of rotation rates of inner and outer cylinders $\mu$ at various helicities
$\gamma$ of rotational helical magnetic field with $\alpha=1,$ $\beta=0$,
and $\mathrm{Ha}=10.$ As the axial current defined by $\gamma$ is
increased, the lower critical Reynolds number reduces and the range
of instability beyond the Rayleigh limit increases until the critical
value $\gamma\approx3.7$ is attained. At this critical helicity,
the lower critical Reynolds number drops to zero and the range of
instability becomes effectively unlimited. It is important to note
that the extension of instability beyond the Rayleigh limit reduces
with the increase of Reynolds number. This corresponds to the restabilization
of the flow by the fast rotation, which takes place above the upper
critical Reynolds number plotted in \ref{fig:rec-gm}(a) for the values
of $\mu$ beyond the Rayleigh limit.

\subsection{Instability in the azimuthal magnetic field generated by axial current
in the liquid}

\begin{figure}
\begin{centering}
\includegraphics[width=0.45\textwidth]{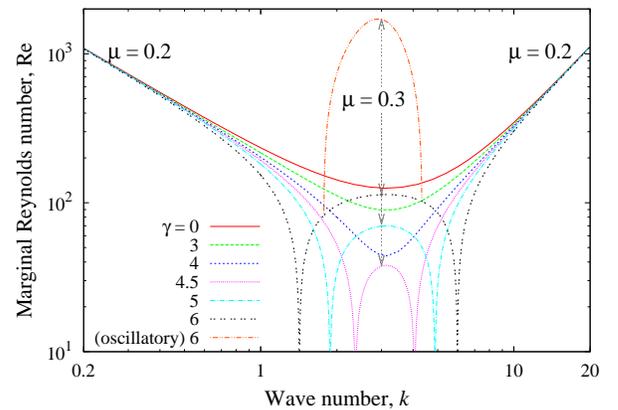}
\par\end{centering}

\protect\caption{\label{fig:rewk-afgm}(Color online) Marginal Reynolds number versus
wave number for hydrodynamically unstable $(\mu=0.2)$ and stable
$(\mu=0.3)$ flows in the azimuthal magnetic field $(\alpha=0)$ generated
only by the axial current in the liquid annulus $(\beta=0)$ with
various magnitude $\gamma$ at $\mathrm{Ha}=10.$ }
\end{figure}

Let us consider next what happens when the axial component of the
magnetic field is switched off by setting $\alpha=0.$ It means that
the magnetic field is now perfectly azimuthal and generated only by
the axial current in the liquid. Marginal Reynolds number and the
frequency for both hydrodynamically unstable $(\mu=0.2)$ and stable
$(\mu=0.3)$ flows in the magnetic fields of various strength defined
by $\gamma$ and $\mathrm{Ha}=10$ are plotted against the wave number
$k$ in Fig. \ref{fig:rewk-afgm}. For the hydrodynamically unstable
flow, the effect of the azimuthal field is very similar to that of
the helical field considered previously. Namely, the increase of the
axial electric current defined by $\gamma$ reduces marginal Reynolds
number, which again drops to zero in a certain range of wave numbers
when $\gamma\gtrsim4.5.$ In contrast to helical magnetic field, now
the instability is completely stationary, i.e., $\omega=0.$ For hydrodynamically
stable flow, the effect slightly differs from that of the helical
field. First, in this case all neutral stability curves, which as
before exist only for a limited range of wave numbers, end at zero
Reynolds number. It means that the lower critical Reynolds number,
if any, is always zero when the flow is hydrodynamically stable. Second,
as seen in Fig. \ref{fig:rewk-afgm}, an oscillatory instability mode
appears contrary to \citet{Edmonds1958} conjecture in a certain subrange
of unstable wave numbers at sufficiently high $\mathrm{Re}$ when
$\gamma\gtrsim6.$ This oscillatory mode, which resembles an electromagnetically
destabilized inertial wave, persists up to much higher Reynolds numbers
than the stationary one. 

The stationary mode looks like a pinch-type instability which has
been studied in this setup numerically by Shalybkov using a more general
non-axisymmetric and finite-$\textit{Pm}$ approximation \citep{Shalybkov2006,Shalybkov2007}
and \citet{Ruediger2007} in the context of the so-called azimuthal
MRI. The latter is inherently nonaxisymmetric \citep{Hollerbach2010}
and has the same limited extension beyond the Rayleigh line as the
HMRI in the highly resistive limit \citep{Kirillov2012}. Undeterred
by the associated identification challenges, which we discussed in
the introduction, \citet{Seilmayer2014} claim to have observed this
instability in another recent liquid-metal experiment.

Pinch-type instability operates through the compression of the azimuthal
magnetic field lines by a radially inward flow perturbation which
amplifies itself by enhancing the electromagnetic pinch force generated
by the interaction of the axial electric current with its own magnetic
field. It is important to notice that axisymmetric meridional flow
interacts only with the free- space $(\sim r^{-1})$ but not with
the rotational $(\sim r)$ component of the azimuthal magnetic field
\citep{Michael1954,Edmonds1958}. As is easy to see from Eq. (\ref{eq:hhat}),
the respective induction term proportional to $\beta-\gamma$ is entirely
due to the free-space component of the magnetic field, and vanishes
together with the latter when $\gamma=\beta.$ The interaction between
axisymmetric meridional flow and azimuthal rotational magnetic field
is precluded by the conservation of the magnetic flux. The flux is
conserved because the rotational magnetic field varies linearly with
the cylindrical radius $r$ while the respective cross-sectional area
of a toroidal element of constant volume in incompressible fluid flow
varies inversely with $r.$ Thus, in contrast to the conventional
$z$ pinch, this instability requires not only a rotational but also
a free-space component of the azimuthal magnetic field. The latter,
however, is possible only in annular but not in cylindrical geometry.
As seen from Eq. (\ref{eq:b-phi}), the free-space component of the
azimuthal magnetic field associated to the axial electric current
in annular geometry $(R_{i}\not=0)$ can be compensated by an additional
free-space magnetic field with $\beta=\gamma$, which leaves only
the rotational component $\sim r$ as in the solid cylinder.

\begin{figure}
\begin{centering}
\includegraphics[width=0.45\textwidth]{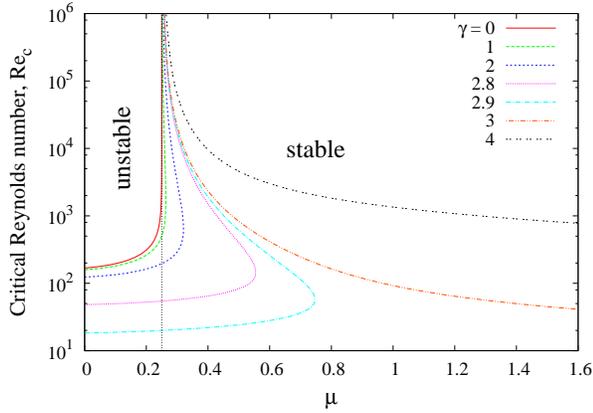}
\par\end{centering}

\protect\caption{\label{fig:rec-btgm}(Color online) Critical Reynolds number versus
the ratio of rotation rates of inner and outer cylinders $\mu$ at
various helicities $\gamma$ of purely rotational helical magnetic
field with $\alpha=1,$ $\beta=\gamma$, and $\mathrm{Ha}=10.$ }
\end{figure}

\subsection{Instability in helical magnetic field \protect \\
with a perfectly rotational azimuthal component}

\begin{figure}
\begin{centering}
\includegraphics[width=0.45\textwidth]{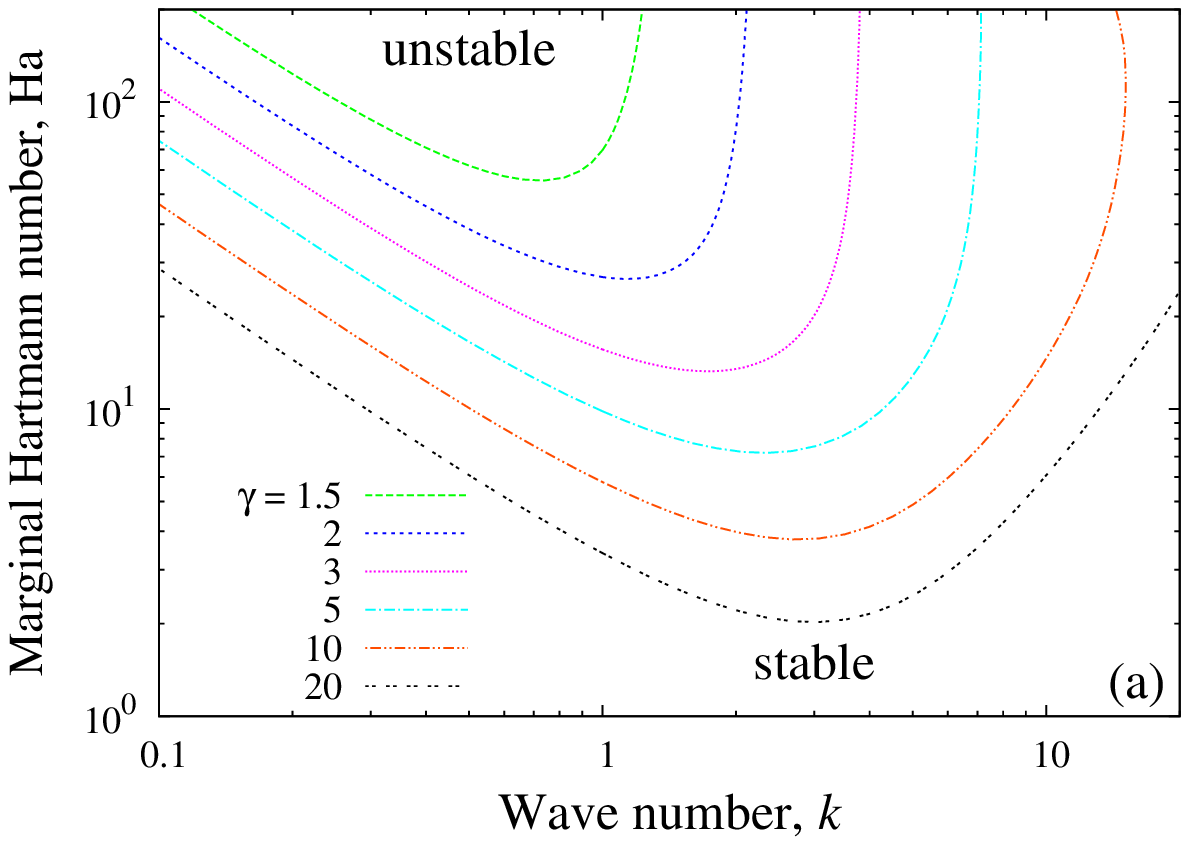}
\par\end{centering}

\begin{centering}
\includegraphics[width=0.45\textwidth]{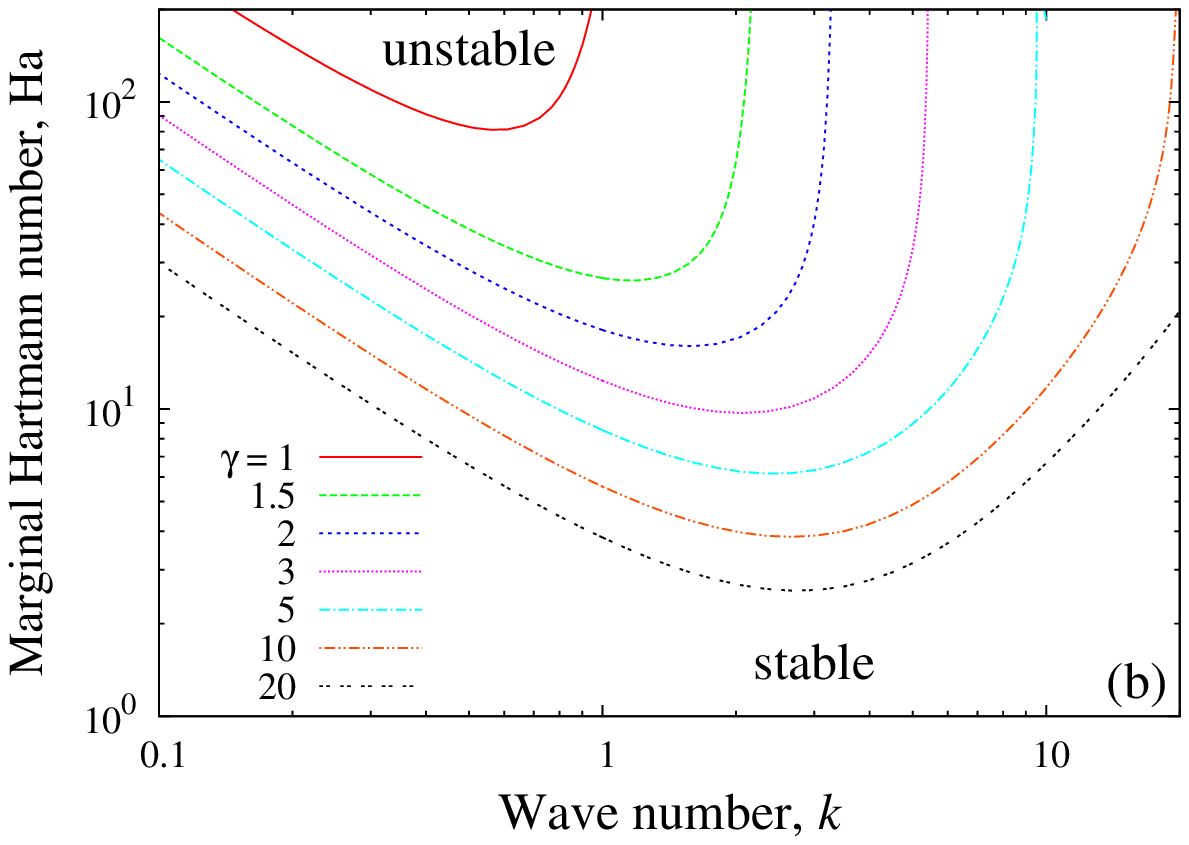}
\par\end{centering}

\begin{centering}
\includegraphics[width=0.45\textwidth]{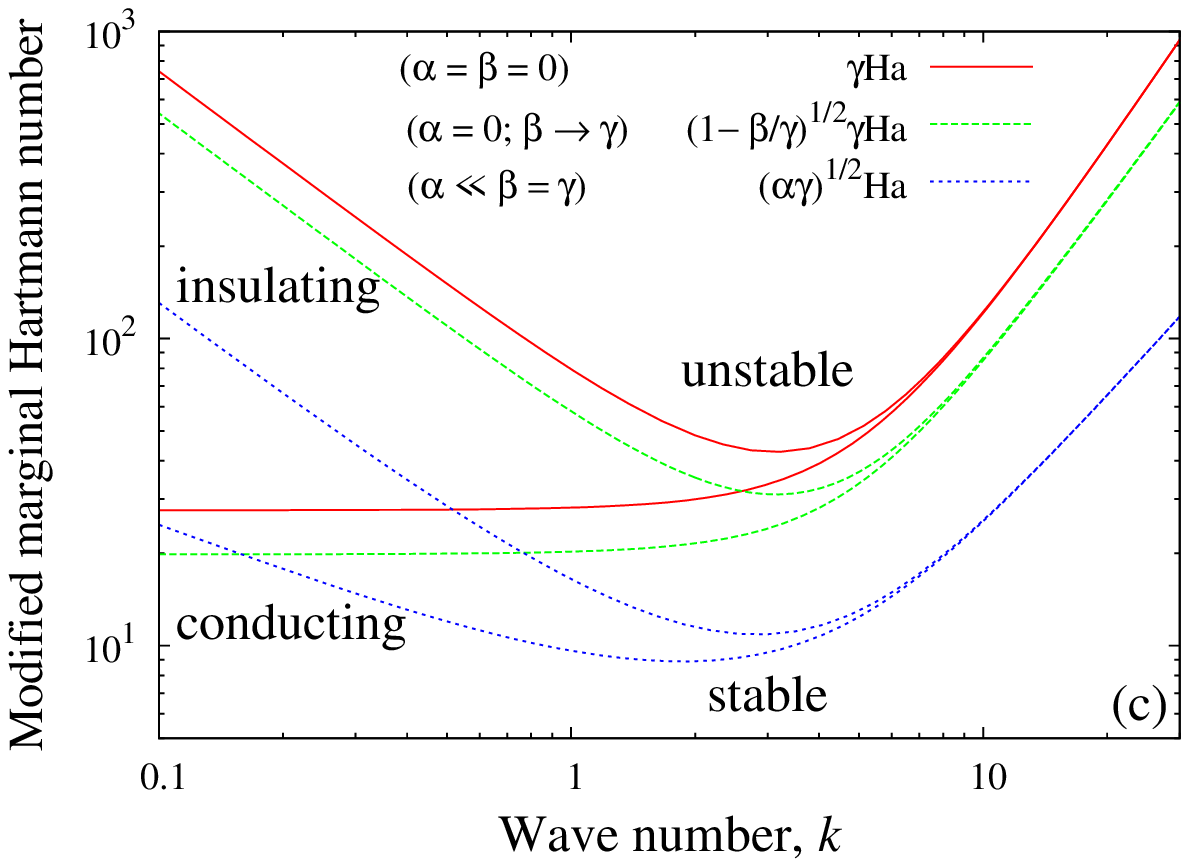}
\par\end{centering}

\protect\caption{\label{fig:Hawk}(Color online) Marginal Hartmann number versus wave
number for purely electromagnetic $(\mathrm{Re}=0)$ stationary $(\omega=0)$
instabilities in the rotational magnetic field with $\alpha=1,$ $\beta=0$
(a), $\alpha=1,$$\beta=\gamma$ (b), and $\alpha=0,$ $\beta=0$,
and $\beta\rightarrow\gamma$ (c) for both insulating and perfectly
conducting cylinders at various axial currents defined by $\gamma.$ }
\end{figure}

Now let us check what happens when the axisymmetric pinch instability
is excluded by applying a compensating free-space magnetic field with
$\beta=\gamma$ which makes the azimuthal component of the magnetic
field perfectly rotational, that is, purely linear in $r.$ In order
to have any electromagnetic effect on the axisymmetric disturbances,
we need to add an axial magnetic field by setting $\alpha=1.$ Both
the critical Reynolds number and the frequency, which are shown in
Fig. \ref{fig:rec-btgm} versus the ratio of rotation rates of outer
and inner cylinders for $\mathrm{Ha}=10,$ look very similar to the
respective characteristics shown in Fig. \ref{fig:rec-gm}(a) for
the rotational helical magnetic field with an uncompensated free-space
component. As before, the increase of the axial current reduces the
critical Reynolds number, which in this case drops to zero at the
critical value $\beta=\gamma\approx2.9$, leading to an unlimited
extension of the instability beyond the Rayleigh limit for larger
values of $\gamma$. Thus, the elimination of the pinch-type instability
has a surprisingly little effect on the remaining instability.

\subsection{Purely electromagnetic instabilities}

\begin{figure}
\begin{centering}
\includegraphics[bb=190bp 90bp 270bp 250bp,clip,height=0.3\textheight]{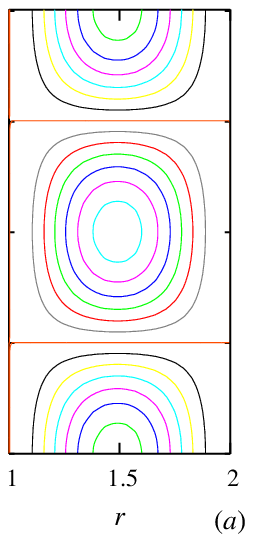}\includegraphics[bb=190bp 90bp 285bp 250bp,clip,height=0.3\textheight]{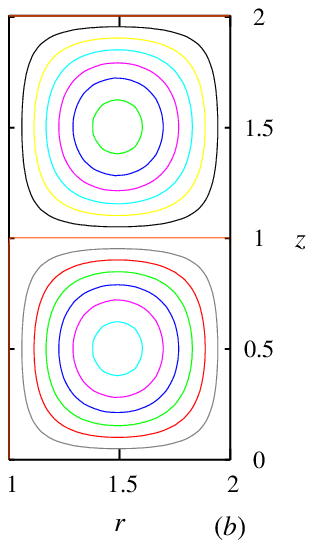}
\par\end{centering}

\protect\caption{\label{fig:eigv-afgm}(Color online) Streamlines (a) and the electric
current lines (a) of the critical perturbation for the electromagnetic
$(\mathrm{Re}=0)$ pinch-type instability $(\alpha=\beta=0)$. }
\end{figure}

Zero marginal Reynolds number means that the instability no longer
depends on the background flow and is driven entirely by the electromagnetic
force which is defined by Hartmann number. Marginal $\mathrm{Ha}$
for such electromagnetically sustained disturbances is plotted in
Fig. \ref{fig:Hawk} against wave number for various axial current
parameters $\gamma$ in helical magnetic field with uncompensated
$(\alpha=1,\beta=0)$ (a) and compensated $\beta=\gamma$ (b) free-space
azimuthal components as well as in a purely azimuthal field $(\alpha=0)$
generated only by the axial current in the liquid $(\beta=0),$ and
with nearly compensated free-space component $(\beta\rightarrow\gamma)$
(c). For the first two helical field configurations, marginal $\mathrm{Ha}$
is seen to vary with $\gamma$ in a similar way. For purely azimuthal
field configuration, the pinch-type instability driven only by the
current passing through the liquid is determined by the effective
Hartmann number $\gamma\mathrm{Ha}.$ As seen in Fig. \ref{fig:Hawk}(c),
the lowest value $\gamma\mathrm{Ha}_{c}\approx42.74$ for insulating
cylinders is attained at the critical wave number $k_{c}\approx3.13$.
When the cylinders are perfectly conducting and thus the induced currents
can freely close through them, the instability threshold is seen to
decrease with the wave length so the lowest value $\gamma\mathrm{Ha}_{c}\approx26.6$
is attained asymptotically at $k\rightarrow0.$ This instability gives
rise to a steady meridional flow whose streamlines and the associated
electric current lines for insulating boundaries are shown in Fig.
\ref{fig:eigv-afgm}. 

\begin{figure*}
\begin{centering}
\includegraphics[width=0.5\textwidth]{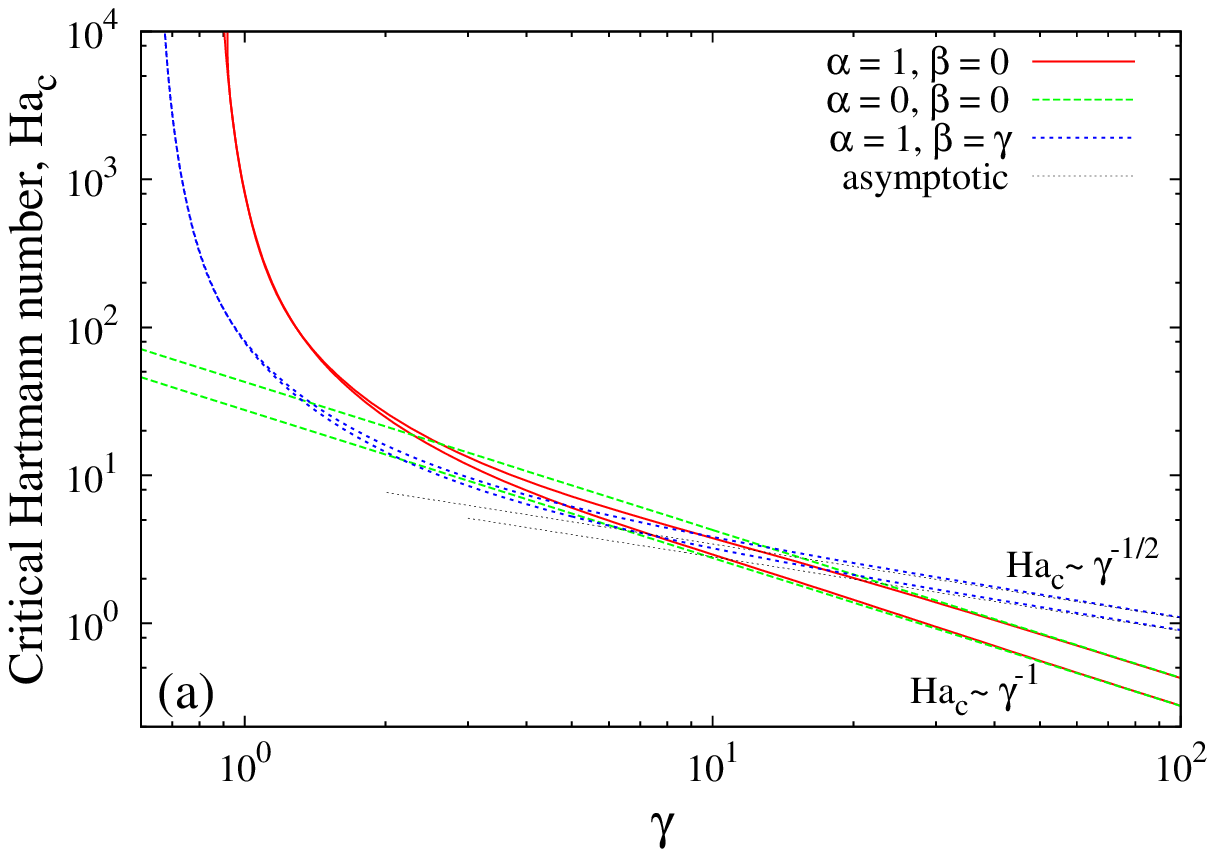}\includegraphics[width=0.5\textwidth]{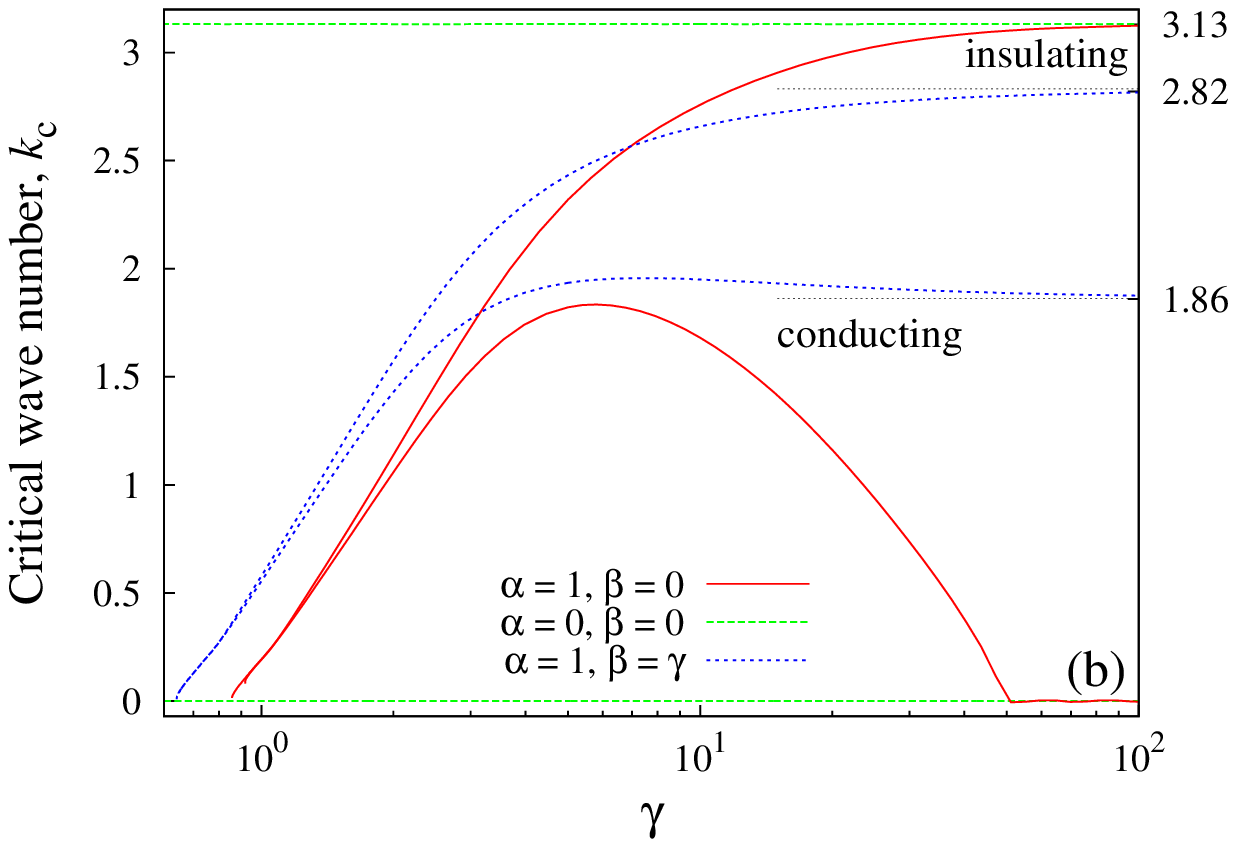}
\par\end{centering}

\protect\caption{\label{fig:Hac-gm}(Color online) Critical Hartmann number (a) and
wave number (b) for purely electromagnetic $(\mathrm{Re}=0)$ stationary
$(\omega=0)$ instabilities versus the axial current parameter $\gamma$
for different magnetic field configurations. The upper and lower branches
for each configuration correspond to insulating and perfectly conducting
cylinders.}
\end{figure*}

\begin{figure*}
\begin{centering}
\includegraphics[bb=195bp 90bp 265bp 250bp,clip,height=0.3\textheight]{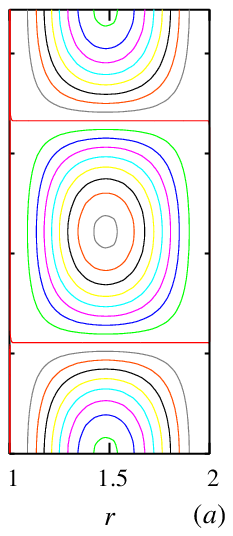}\includegraphics[bb=195bp 90bp 265bp 250bp,clip,height=0.3\textheight]{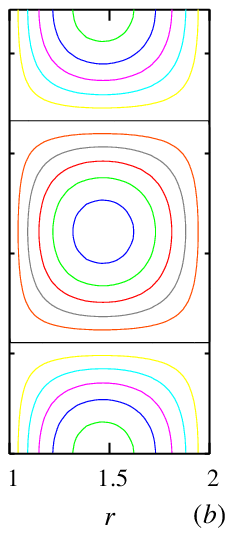}\includegraphics[bb=195bp 90bp 265bp 250bp,clip,height=0.3\textheight]{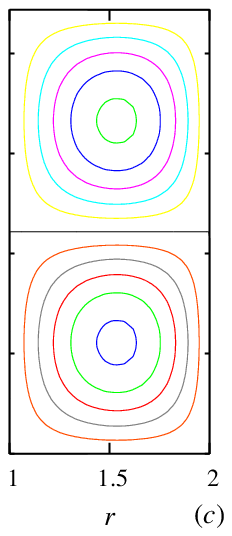}\includegraphics[bb=165bp 90bp 320bp 250bp,clip,height=0.3\textheight]{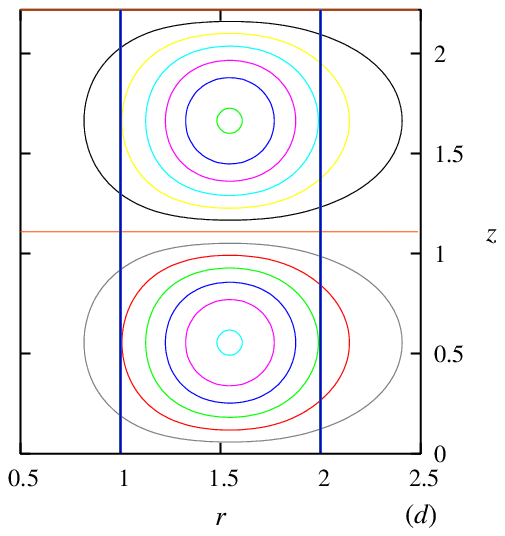}
\par\end{centering}

\protect\caption{\label{fig:eigv-afbtgm}(Color online) Streamlines (a), isolines of
the azimuthal velocity (b), the electric current lines (c), and the
meridional magnetic flux lines (d) for the critical perturbation of
electromagnetic $(\mathrm{Re}=0)$ rotational instability in the asymptotic
case $\alpha\ll\beta=\gamma$. }
\end{figure*}

Critical Hartmann and wave numbers for all three basic field configurations
are summarized in Fig. \ref{fig:Hac-gm} for both insulating and perfectly
conducting boundaries. It is seen that at a sufficiently large $\gamma,$
the instability in helical magnetic field with a nonzero (uncompensated)
free-space azimuthal component turns into the pinch instability with
$\mathrm{Ha}_{c}\sim\gamma^{-1}.$ When this pinch-type instability
is excluded by setting $\beta=\gamma,$ which corresponds to a compensated
free-space azimuthal component of the magnetic field, the critical
Hartmann number at large $\gamma$ varies differently as $\mathrm{Ha}_{c}\sim\gamma^{-1/2}.$
This implies a different type of instability which is driven by the
interaction of axial electric current with a collinear external magnetic
field. As seen in Fig. \ref{fig:Hawk}(c), in contrast to the pinch-type
instability, this instability has a finite critical wave length not
only for insulating but also for perfectly conducting cylinders. The
latter fact implies that this instability relies on the closure of
induced currents within the liquid which is discussed in more detail
in the concluding section. The critical perturbation pattern of this
rather complex instability for insulating cylinders is shown in Fig.
\ref{fig:eigv-afbtgm}. 

\begin{figure}
\centering{}\includegraphics[width=0.45\textwidth]{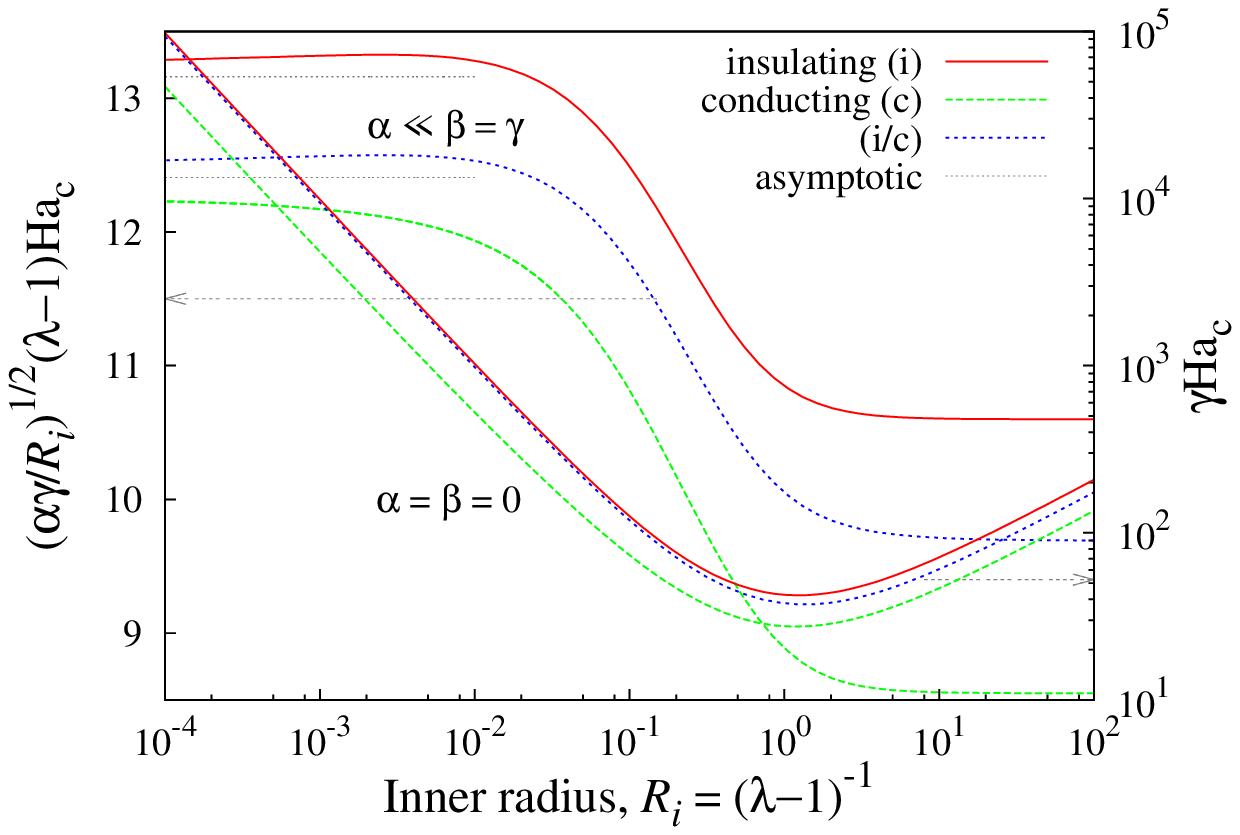}\protect\caption{\label{fig:Hac-Ri}(Color online) Critical Hartmann number rescaled
with the gap width $\lambda-1$ versus the rescaled inner radius $R_{i}=(\lambda-1)^{-1}$
for the pinch instability $(\alpha=\beta=0)$ and for the instability
driven by the electric current in a weak axial magnetic field ($\alpha\ll\beta=\gamma$)
with both cylinders insulating (i), perfectly conducting (c), inner
cylinder insulating and outer cylinder perfectly conducting (i/c).
The Hartmann number for the latter instability (on the left axis)
is additionally rescaled with the effective current density $\gamma/R_{i}.$}
\end{figure}

Figure \ref{fig:Hac-Ri} shows the critical Hartmann number based
on the gap width against the inner radius of the annular gap for both
the pinch-type instability $(\alpha=\beta=0)$ and the instability
driven by the electric current in a weak axial magnetic field ($\alpha\ll\beta=\gamma$).
In the limits $R_{i}\rightarrow0$ and $R_{i}\rightarrow\infty,$
the annular layer turns into a cylinder and a flat layer, respectively.
Note that according to our parametrization of the magnetic field (\ref{eq:b-phi}),
the axial current density diverges as $\gamma/R_{i}\rightarrow\infty$
when $R{}_{i}\rightarrow0.$ The critical Hartmann number for the
pinch-type instability based on this singular current density approaches
a constant value when $R{}_{i}\rightarrow0$ and to increases as $\sim R_{i}^{3/2}$
for $R_{i}\gg1.$ The critical Hartmann number based on the fixed
current density, i.e., rescaled with $R_{i}^{-1},$ which is plotted
in Fig. \ref{fig:Hac-Ri}, attains minima at $R_{i}\approx1$ and
increases as $\sim R_{i}^{-1}$ and $\sim R_{i}^{1/2}$ for $R_{i}\ll1$
and $\gg1,$ respectively. It means that, the related pinch instability
vanishes not only in the cylindrical geometry, where $\beta=\gamma,$
but also in the planar unbounded layer, where the associated electromagnetic
force can be shown become purely irrotational. The critical Hartmann
number rescaled with the current density for the other instability
is seen to remain finite in both limits of $R_{i}.$ It means that
in contrast to the pinch instability, this instability has a finite
critical current density also in the cylindrical geometry. However,
despite the finite critical Hartmann number for $R_{i}\rightarrow\infty,$
this instability has no analog in the planar unbounded layer. In this
case, critical Hartmann number remains finite when $R_{i}\rightarrow\infty$
because the azimuthal magnetic field (\ref{eq:b-phi}) for $\beta=\gamma$
diverges as $\sim R_{i}.$ When the current density is rescaled by
$1/R_{i}$ to have a finite magnetic field when $R_{i}\rightarrow\infty,$
the critical Hartmann number for $R_{i}\gg1$ also increases as $\sim R_{i}^{1/2}.$

\begin{figure}
\begin{centering}
\includegraphics[width=0.45\textwidth]{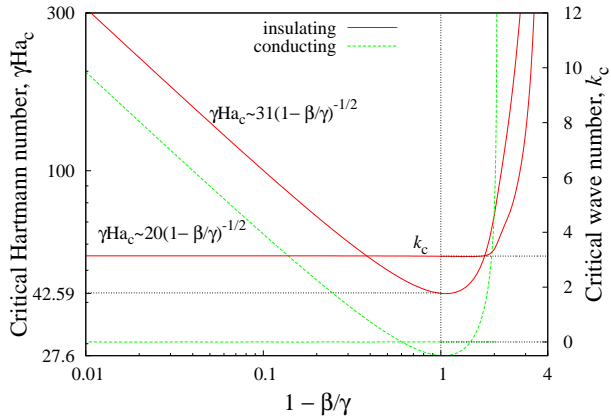}
\par\end{centering}

\protect\caption{\label{fig:Hac-afbt}(Color online) Critical Hartmann number and the
wave number for the electromagnetic pinch instability versus $1-\beta/\gamma$
which defines the strength of free-space azimuthal component the magnetic
field relative to the rotational component generated by axial current. }
\end{figure}

Finally, let us consider the effect of additional free-space azimuthal
field on the pinch-type instability without axial magnetic field $(\alpha=0).$
Critical Hartmann number and the wave numbers for this case are shown
in Fig. \ref{fig:Hac-afbt} versus $1-\beta/\gamma$, which defines
the relative strength of the free-space component. As seen in Fig.
\ref{fig:Hac-afbt}, the critical Hartmann number attains the minimum
$\mathrm{Ha}_{c}\approx42.74\gamma^{-1}$ at $\beta/\gamma$ slightly
less than $0$ and increases asymptotically as $\mathrm{Ha}_{c}\sim31\gamma^{-1}(1-\beta/\gamma)^{-1/2}$
when $\beta/\gamma\rightarrow1.$ Marginal Hartmann number for this
limit, which corresponds to a nearly compensated free-space azimuthal
component of the magnetic field, is plotted in Fig. \ref{fig:Hawk}(c).
Asymptotic result is obtained by dropping the term with $\beta-\gamma$
in Eq. (\ref{eq:omghat}) which produces a quadratically small effect
relative to analogous term in Eq. (\ref{eq:omghat}) when $\beta\rightarrow\gamma.$
Critical Hartmann number becomes very large also when $\beta/\gamma\rightarrow-2,$
which corresponds a compensated total axial current through the system.
In this case, the current which passes through the liquid returns
along a central electrode and thus cancels the field in the free space
outside the system. This setup has been suggested by \citet{Stefani2011}
as a possible means of avoiding pinch-type instability in the future
liquid metal batteries.

\section{\label{sec:end}Summary and conclusions}

The present study was concerned with numerical linear stability analysis
of a cylindrical Taylor-Couette flow of liquid metal carrying an axial
electric current in the presence of a generally helical external magnetic
field. It was shown that the electric current passing through the
liquid profoundly alters the nature of the helical MRI by transforming
it into a purely electromagnetic instability. Two different electromagnetic
instability mechanisms were identified. The first is an internal pinch-type
instability which is driven by the interaction of the electric current
with its own magnetic field. The axisymmetric mode of this instability
considered in the present study requires a free-space component of
the azimuthal magnetic field, which is possible in the annular but
not in the cylindrical geometry. In the annular geometry this instability
mode can be eliminated by passing an additional current along the
axis of the system to compensate the free-space azimuthal component
of the magnetic field in the liquid. In this case, the addition of
axial magnetic field was found to give rise to a new kind of electromagnetic
instability. 

The mechanism of this instability, which is driven by the interaction
of axial electric current with a weak collinear external magnetic
field, is as follows. First, a radially outward initial flow perturbation
slightly bends the axial magnetic field but does not affect, as argued
above, the purely rotational azimuthal field and the associated axial
current. The deflected axial field crossing the unperturbed axial
current gives rise to an azimuthal electromagnetic force which, in
turn, drives an azimuthal flow perturbation. The fluid rotates in
the positive direction below the radial flow perturbation, where the
axial field is bent outwards, and in the negative direction above
it, where the axial field bends back. Next, the azimuthal flow perturbation
in the axial magnetic field induces radially outward and inward electric
currents below and above the initial radial flow perturbation, respectively.
These two opposite radial electric currents close in the inner part
of the liquid annulus via a downward axial current, which, in turn,
interacts with the azimuthal magnetic field and generates a radially
outward electromagnetic force perturbation. The latter amplifies the
initial radial flow perturbation, so promoting the instability. 

In contrast to the azimuthal MRI \citep{Ruediger2007}, the helical
MRI does not separate from purely electromagnetic instabilities in
the inductionless limit $\mathrm{Pm}=0.$ It is also important to
note that although electromagnetic instabilities can develop without
mechanical rotation, the latter has a stabilizing effect when the
base flow is hydrodynamically stable. Similarly to the HMRI, the electromagnetic
instabilities are constrained beyond the Rayleigh line to sufficiently
low Reynolds numbers. This dynamical constraint may severely limit
astrophysical relevance of electromagnetic instabilities. Nevertheless,
there are several industrial applications such as, for example, aluminum
reductions cells \citep{Pedchenko2009} and the prospective liquid
metal batteries \citep{Wang2014}, where the strong electric current
passing through the liquid metal in the presence of a collinear magnetic
field can give rise to the electromagnetic instability identified
in this study. 
\begin{acknowledgments}
This work was supported by Helmholtz Association of German Research
Centres (HGF) in the framework of the LIMTECH Alliance through an
agreement between Coventry University and Helmholz-Zentrum Dresden-Rossendorf.
\end{acknowledgments}

\bibliography{pinch_rev}

\end{document}